\documentstyle[11pt]{elsart}

\begin{document}
\begin{frontmatter}
 \title{Fast and Flexible CCD Driver System Using Fast DAC and FPGA}

 \author[Osaka,CREST]{Emi Miyata},
 \author[Osaka]{Chikara Natsukari},
 \author[Osaka]{Daisuke Akutsu},
 \author[Osaka]{Tomoyuki Kamazuka}
 \author[genshi]{Masaharu Nomachi},  and
 \author[isas]{Masanobu Ozaki}

 \address[Osaka]{Department of Earth \& Space Science, Graduate School 
 of Science, Osaka University, 1-1 Machikaneyama, Toyonaka, Osaka
 560-0043, Japan}
 \address[genshi]{Department of Physics, Graduate School 
 of Science, Osaka University, 1-1 Machikaneyama, Toyonaka, Osaka
 560-0043, Japan}
 \address[CREST]{CREST, Japan Science and Technology Corporation (JST)}
 \address[isas]{3-1-1 Yoshinodai, Sagamihara Kanagawa 229-8510, Japan}

 \begin{abstract}
  We have developed a completely new type of general-purpose CCD data
  acquisition system which enables one to drive any type of CCD using
  any type of clocking mode.  A CCD driver system widely used before
  consisted of an analog multiplexer (MPX), a digital-to-analog
  converter (DAC), and an operational amplifier. A DAC is used to
  determine high and low voltage levels and the MPX selects each voltage
  level using a TTL clock. In this kind of driver board, it is difficult
  to reduce the noise caused by a short of high and low level in MPX and
  also to select many kinds of different voltage levels.

  Recent developments in semiconductor IC enable us to use a very fast
  sampling ($\sim$ 10MHz) DAC with low cost. We thus develop the new
  driver system using a fast DAC in order to determine both the
  voltage level of the clock and the clocking timing. We use FPGA
  (Field Programmable Gate Array) to control the DAC. We have constructed
  the data acquisition system and found that the CCD functions well with our 
  new system. The energy resolution of Mn K$\alpha$ has a full-width at
  half-maximum of $\simeq$ 150 eV and the readout noise of our system is
  $\simeq$ 8 e$^-$.
 \end{abstract}

\end{frontmatter}

 \section{Introduction}

 Most recent X-ray satellites carry a charge-coupled device (CCD)
 camera for their focal plane instrument. CCD's possesses a moderate
 energy resolution, a high spatial resolution, and a timing resolution.
 The Solid-state Imaging Spectrometer, SIS, onboard ASCA was the first
 CCD camera used as a photon counting detector. 
 (Tanaka et al\cite{Tanaka}). Following the SIS, many satellites
 such as HETE2 (Ricker\cite{Ricker}), Chandra (Weisskoph et
 al.\cite{ACIS}), XMM-Newton (Barr et al.\cite{XMM}), and MAXI (Matsuoka
 et al.\cite{Matsuoka}) now carry a X-ray CCD camera on their focal planes.

 MAXI, Monitor of All-sky X-ray Image, has been selected as an early
 payload of the JEM (Japanese Experiment Module) Exposed Facility on the
 International Space Station. MAXI will monitor the activities of about
 2000-3000 X-ray sources. It consists of two kinds of X-ray detectors:
 the first, the gas slit camera (GSC) is a one-dimensional 
 position-sensitive proportional counter, and the other, the
 solid-state slit camera (SSC) is an X-ray CCD array.
 The CCD used in the SSC is fabricated by
 Hamamatsu Photonics K.K. (HPK) and is being calibrated both at Osaka
 University and the National Space Development Agency of Japan (NASDA).

 Since SSC is the first CCD camera fabricated soley by Japan, we need
 to specify the functioning of the CCD in detail. In order to optimize the
 function of the CCD, we need to develop a highly flexible data acquisition
 system.

 \section{Requirements for New System}\label{sec:requirements}

 In order to optimize the X-ray responsibility of the CCD, we need to
 develop a highly flexible CCD driver. Our requirements of the CCD
 driver are:

 \begin{itemize}
  \item to output any kind of clocking pattern
  \item to dynamically control clocking voltages
  \item to modify the clocking pattern easily and download it by request
  \item to have a readout speed $\ge$ 1MHz
  \item to output clocking voltages with a range of $-20$ to $+20$ V
  \item to control voltage levels to within 0.1 V
 \end{itemize}

 The clock driver circuit used until now consists of MPXs,
 DACs (digital-to-analog converters), and analog
 amplifiers. For example, two DACs are used to generate the low and high
 voltage level of a clock and an MPX switches each level
 with a digital signal. This system has been well established but it is
 not suitable to change the voltage level dynamically.

 \section{New CCD Data Acquisition System}

 To satisfy all of the requirements listed in section~\ref{sec:requirements},
 we have developed a new type of CCD driver system as shown in
 figure~\ref{fig:sche}. We use one fast DAC to generate each clock. This
 enables us to control each clock with a high flexibility whereas we
 need a lot of control I/O pins. In the previous system, the voltage
 level of each DAC is determined before operating the CCD and at least
 one I/O pin is needed for each clock. On the other hand, our new system
 requires the number of clocks times 10 pins per DAC even if we use an 8-bit
 DAC, resulting in roughly orders of magnitude more I/O pins than the previous
 system.  We thus introduced a field programmable gate array (FPGA) to
 control all DACs.

 \subsection{Design of the DAC board}
 
 Because a CCD is operated by DACs directly, the noise characteristics
 need to be low.  We therefore picked up more than five DACs to evaluate
 the noise characteristics.  Among them, TLC 7524 fabricated by 
 Texas Instruments possesses the lowest noise characteristics and we
 select this device for our new system.

 A detailed design around DAC in figure~\ref{fig:sche} is shown in
 figure~\ref{fig:dac}. We use a photo-coupler, HCPL$-$2430, to separate
 an analog and a digital ground. TLC7524 is an 8-bit current-output DAC
 whose settling time is $\sim$ 100 ns. The fast settling time enables us
 to simultaneously control both the clocking timing and the voltage
 level, which is realized with several DACs and MPXs in the previous
 driver system. Thus, our new system posses a high flexibility though it
 is much simpler than the previous system.

 \subsection{Design of the FPGA board}

 We previously used the VME system to control the DAC boards and had a
 lot of noise problems mainly due to a switching regulators on a VME
 power supply unit. We thus abandon using the VME system for this
 purpose.  We designed a general-purpose digital I/O board (DIO board)
 to simultaneously control several DAC boards.  Our DIO board carries a
 reconfigurable FPGA, 512 Kbyte SRAM device (PD434008ALE-15), a serial
 interface, a parallel interface with 10 bits, an interface for a liquid
 crystal to display the status, and eight DAC interfaces. One DAC
 interface possesses 10 bits in order to control a 10 bit DAC in a
 future application.  Figure~\ref{fig:fpga_board} shows a photograph of
 the FPGA board developed in this work. We selected an Altera Flex 10K50
 for the FPGA. This FPGA device is a static memory type that can be
 reconfigured simply with a command and has 189 pins available for the
 user.

 One of the remarkable advantages is the development of 
 Hardware Description Language (HDL). HDLs and synthesis tools can
 greatly reduce the design time, improving the time-to-market. A description
 based on HDLs is easier to understand than some schematic for a very
 large design in FPGA gate format. There are several kinds of HDLs
 developed for various corporations: AHDL\cite{ahdl}, VHDL\cite{vhdl},
 and Verilog-HDL\cite{verilog}. Among them, we employed VHDL.
 Throughout the development, we used the MAX+PLUS II and FPGA Express
 software provided by Altera corporation and Synopsis corporation.

 \subsection{Data Acquiring System}

 The CCD output signal is processed with a delay line and peak-hold
 circuits which have been previously developed by our group. The
 processed signal is shifted to $\pm$ 5V and sampled by a 12-bit
 analog-digital converter (ADC). Digital data are transferred to the VME
 I/O board (\cite{kataoka}) with a flat cable and are sent to the
 sparcstation through the VME bus.

  \subsection{Sequencer}

  We have developed a sequencer and relevant software to compile it. We
  define two sequencers: V-ram and P-ram. The V-ram defines a voltage
  pattern to drive the CCD with a relatively a short duration. Combining 
  several V-rams, we describe the clocking pattern for readout of whole CCD 
  in P-ram.

   \subsubsection{V-ram}

   We develop, typically, two kinds of V-ram: V-ram for readout one
   pixel and transfer one line. An example of V-ram for one pixel
   readout in a two phase CCD is shown below.  The vertical axis represents
   the time sequence. {\tt P1H} and {\tt P2H} are clocks for the serial
   register and {\tt P1V} and {\tt P2V} are those for the vertical
   register. {\tt RST} and {\tt HOLD} are clocks for reset and
   ADC. Numbers in V-ram represent the voltage level in units of
   Volts. Following brackets show that a voltage level is the same as the
   previous value. In this way, we describe the voltage level and the
   timing for a voltage change in V-ram.

   The V-ram compiler we developed reads the V-ram and creates the DAC
   patterns for each clock. The {\tt HOLD} signal is transferred to the ADC 
   board through the parallel interface while others are transferred to
   the appropriate DAC interace.

   In the current system, we use TLC7524 which needs a reference
   clock. When the reference clock is sent to TLC7524, it latches all
   data bits and outputs the voltage depending on the data bits. Since
   the reference clock is difficult to be described in V-ram, the V-ram
   compiler adds it in the output sequencer code automatically.

   \begin{verbatim}
                   P1H     P2H     RST     P1V     P2V    HOLD
                   -8       6      6       6       6       5 
                   [        ]       ]       ]       ]       ]
                   [        ]      -8       ]       ]       ]
                   [        ]      [        ]       ]       ]
                   [        ]      [        ]       ]       ]
                   [        ]      [        ]       ]       ]
                   [        ]      [        ]       ]       ]
                   [        ]      [        ]       ]       ]
                   6       -8      [        ]       ]      0
                    ]      [       [        ]       ]      [
                    ]      [       [        ]       ]      [
                    ]      [       [        ]       ]      [
                    ]      [       [        ]       ]      [
                    ]      [       [        ]       ]      [
                    ]      [       [        ]       ]      [
                    ]      [       [        ]       ]      [
   \end{verbatim}

   \subsubsection{P-ram}

   P-ram is described to define the readout of a whole CCD.  To include
   V-ram files, P-ram uses the filename of V-ram. We have prepared
   several instruction commands to describe any P-rams easily and
   concisely as listed in table~\ref{table:pram_command}.  Combining
   filenames of V-rams and instruction commands, P-ram can be easily
   developed by the user. One example of a P-ram is shown below.

    \begin{verbatim}
                  set A = 64
                  set B = 2
                  set xaxis = 1024
                  set yaxis = 1024

                  start:
                   	do yaxis
                        	set wait A
                        	seq 1 vertical
                        	set wait B
                        	seq xaxis horizontal
                   	end do
                  jmp start
    \end{verbatim}

    This P-ram reads out a CCD with 1024 $\times$ 1024 pixels.
    V-ram of 'vertical' and 'horizontal' define the voltage pattern to
    transfer pixels vertically and horizontally, respectively.  The
    instruction of 'set wait' is to determine the duration of each level
    in S-ram. 

    The P-ram is compiled on a SUN sparcstation (Force, CPU-50GT) and
    stored in the memory of VME I/O board. After sending a command from
    the sun, P-ram is downloaded to the DIO board by means of the serial
    interface and stored in the memory of the DIO board.

    \subsection{Configuration of the Circuit in FPGA}

    To realize the function of the sequencers, we divided the
    configuration of the FLEX device into five blocks as shown in
    figure~\ref{fig:block_diagram}. Each block is constructed by a
    synchronous state machine. The {\tt Serial Interface} is the
    interface to the VME I/O board to download sequencers. After loading
    sequencers, the {\tt Serial Interface} sends a trigger signal to the
    {\tt Clock Controller}. The state machine of the {\tt Clock
    Controller} is shown in figure~\ref{fig:state_machine}.  The {\tt
    Clock Controller} is in the {\it idle} state until a trigger signal
    is sent. Once the trigger signal is received, the {\tt Clock
    Controller} moves to the {\it memory check} state, where the {\tt
    Clock Controller} sends the memory address and a trigger signal to
    the {\tt Synthesize Pattern}. The {\tt Synthesize Pattern} sends a
    trigger signal to the {\tt Memory Controller} and receives memory
    data. After repeating three times, the {\tt Synthesize Pattern}
    arranges the data into 96 bits and sends it to the {\tt Clock
    Controller}.  Then, the state moves to {\it fetch} where 96 bit data
    is stored in a register and next moves to {\it decode}. In the {\it
    decode} state, the {\tt Clock Controller} analyzes the bit pattern based
    on the instruction commands shown in table~\ref{table:pram_command}
    and sends DAC patterns to the appropriate DAC interface.  After
    sending the DAC patterns, the memory pointer is incremented and {\tt
    Clock Controller} waits for the wait parameter ({\tt A} or {\tt B}
    shown in P-ram). In each state, the {\tt Clock Controller} sends 
    status information to the {\tt Display Controller} and the {\tt
    Display Controller} controls the liquid crystal to display the
    clocking status.
    
    \section{Performance}

    \subsection{Driver System}
    
    In order to demonstrate the performance of our new CCD driver, we
    produced 5 value clockings as shown in figure~\ref{fig:demo}. This
    kind of multi-level clocking is efficient to reduce the spurious
    charge\cite{janesick}.  We thus confirm the high potential and high
    flexibility for our new system.

    Since we use 8 bits DAC for each clock, we can control a voltage
    level within $\simeq$ 0.1 V. We normaly operate the DAC boards with
    ranges of $-$15 to $+$15V. If we change the resister of R13, R15,
    and R20 in figure~\ref{fig:dac}, we can output the clock up to
    $+$20V or down to $-$20V.

    The readout speed is limited by the number of state machines to read
    a voltage pattern from S-ram. In our current design, there are 13
    steps to fetch a 96 bits voltage pattern, resulting in the maximum
    clocking speed to be $\simeq$ 300 KHz. We still need to optimize it
    in order to meet our requirements ($\sim$ 1 MHz).

    \subsection{Total System}

    We compared the performance of our new system with the HPK C4880
    system, which is an X-ray CCD data acquisition system previously
    used\cite{miyata}. We used a CCD chip fabricated by HPK. We cooled
    the CCD down to $-100^\circ$C and irradiated it with an $^{55}$Fe
    source. For comparison, we set the same readout speed as that of
    C4880 (50 KHz). We selected the ASCA grade 0 events\cite{ayamashi}
    with a split threshold of 90 eV and fitted the histogram with two
    Gaussian functions for Mn K$\alpha$ and K$\beta$. Results are shown
    in table~\ref{table:comparison}. The readout noise of our new system
    is $\simeq$ 8e$^-$. We can confirm that our new system function much
    better than the previous system.

    \section{Summary and Future Developments}

    We have developed a new type of general-purpose CCD data acquisition
    system which enables us to drive any kind of CCD with any kind of
    clocking and voltage patterns. It functions well and demonstrates
    great flexibility. We found the readout noise of the CCD to be 8
    e$^-$ rms in our system, which might be contributed to by our
    readout circuit rather than a CCD chip itself.

    We plan to develop the analog electronics to process a CCD output
    signal to reduce the readout noise.  The system currently used is a
    delay line circuit which has poorer performance than an integrated
    type circuit for both readout speed and for noise characteristics
    (especially high frequency regime). Therefore, we will develop an
    integrated correlated double sampling circuit in the near future.

    We also plan to replace the VME I/O board with another FPGA board
    which has already been constructed by us. On this board, 80M
    sampling ADC, FLEX 10K and 4Mbyte S-ram are mounted. There are three
    IEEE 1394 ports each of which has a capability of 400 Mbps
    connection.  Large amount of memory gives us to extract X-ray events
    before sending raw frame data to the host machine. Since FPGA has a
    good capability of a parallel processing comparing with DSP or CPU,
    it enables us to analyze data in real-time. It is also important to
    develop onboard digital processing software using HDL for future
    X-ray astronomy missions.

    We will calibrate the CCD for the MAXI mission with our system. We
    need to determine the voltage pattern and the voltage level to
    optimize the X-ray responsibility.

  \vspace{2cm}
  
  We wish to thank Prof. H. Tsunemi for his valuable comments on the
  initial phase of this work.  We acknowledge to Mr. C. Baluta for his
  critical reading of the manuscript.  This research is partially
  supported by ACT-JST Program, Japan Science and Technology
  Corporation.\\

\clearpage

   \begin{table}[h]
    \begin{center}
     \caption{Instruction commands for P-ram}
     \label{table:pram_command}
     \begin{tabular}{lll}\hline \hline
      Command & Arguments & Function \\
      \hline
      {\tt start} & --- & named label \\
      {\tt jmp} & {\tt label} & jump to  {\tt label} \\
      {\tt seq} & {\tt number, V-ram name} & output {\tt V-ram name} with {\tt
      number} times \\
      {\tt set wait} & {\tt number} & define the output speed \\
      {\tt do} & {\tt number} & repeat all V-rams before next {\tt end do}
      {\tt number} times \\
      {\tt end do} & --- & define the end of block to be repeated \\
      {\tt \#} & --- & write comment \\
      \hline
     \end{tabular}
    \end{center}
   \end{table}

\clearpage

  \begin{table}[h]
   \begin{center}
    \caption{Comparison of our new system with the HPK C4880 system}
    \label{table:comparison}
    \begin{tabular}{lll}\hline \hline
     & C4880 & New system \\ \hline
     Energy resolution [eV] & 162$\pm$3  & 150$\pm$3  \\
     Dark current [e$^-$/pixel/sec] & 0.20$\pm$0.15 & 0.20$\pm$0.14  \\
     Readout noise [e$^-$ rms] & 8.6$\pm$0.5 & 8.0$\pm$0.5 \\
     Exposure time [sec] & 8 & 8 \\
     \hline
    \end{tabular}
   \end{center}
  \end{table}

\clearpage

 \begin{figure}[htbp]
  \caption{The block diagram of the CCD signal flow.}
  \label{fig:sche}
 \end{figure}


 \begin{figure}[htbp]
  \caption{A circuit diagram of the DAC board of figure~\ref{fig:sche}}
  \label{fig:dac}
 \end{figure}


 \begin{figure}[htbp]
  \caption{The picture of the VME I/O board.
  FPGA is mounted around the center of
  the board.}
  \label{fig:fpga_board}
 \end{figure}


  \begin{figure}[htbp]
   \caption{The block diagram of the DIO board. Five gray-colored boxes
   represents the circuits designed in the FLEX chip. There are eight
   DAC interfaces each of which has 10 bits to control the DAC board.}
   \label{fig:block_diagram}
  \end{figure}


  \begin{figure}[htbp]
   \caption{State machine of {\tt Clock Controller} in FPGA}
   \label{fig:state_machine}
  \end{figure}


 \begin{figure}[htbp]
  \caption{Sample clock of multiple levels}
  \label{fig:demo}
 \end{figure}


  \begin{figure}[htbp]
   \caption{Single event $^{55}$Fe spectrum obtained with our new system.}
   \label{fig:fe}
  \end{figure}

\end{document}